\documentclass{aa}  

\usepackage{graphicx}
\usepackage{txfonts}
\begin{document}
   \title{Power Spectrum Analysis of Far-IR Background Fluctuations 
in 160 $\mu$m Maps From the Multiband Imaging Photometer for Spitzer}

          \author{B. Grossan
          \inst{1,2}
          \and G. F. Smoot\inst{3,4}}          
    \institute{Eureka Scientific, Inc.,
    	2452 Delmer Street Suite 100, Oakland, CA 94602-3017
	\and
	 Institute for Nuclear and Particle Astrophysics, 50R-5008, Lawrence Berkeley National Laboratory, 1 Cyclotron Road, Berkeley, CA 94720-8158\\
              \email{Bruce\_Grossan@lbl.gov}
         \and
	University of California at Berkeley, Department of Physics, LeConte Hall, Berkeley, CA 94720
	\and
	50-5005, Lawrence Berkeley National Laboratory, 1 Cyclotron Road, Berkeley, CA 94720-8158\\
             \email{GFSmoot@lbl.gov} }

\offprints{B. Grossan, Bruce\_Grossan@lbl.gov}
 
   \date{Received 2006 April 24 / Accepted 2007 August 29}

   \abstract{We describe data reduction and analysis of fluctuations
     in the cosmic far-IR background (CFIB) in observations with the
     Multiband Imaging Photometer for Spitzer (MIPS) instrument 160
     \ensuremath{\mu}m detectors.  We analyzed observations of an 8.5
     square degree region in the Lockman Hole, part of the largest
     low-cirrus mapping observation with this instrument.  We measured
     the power spectrum of the CFIB in these observations by fitting a
     power law to the IR cirrus component, the dominant foreground
     contaminant, and subtracting this cirrus signal.  The CFIB power
     spectrum in the range $0.2$ arc min$^{-1} <$ \textit{k} $< 0.5$
     arc min$^{-1}$ is consistent with previous measurements of a
     relatively flat component.  However, we find a large power excess
     at low \textit{k}, which falls steeply to the flat component in
     the range $0.03$ arc min$^{-1} <$ \textit{k} $< 0.1$ arc
     min$^{-1}$. This low-\textit{k} power spectrum excess is 
     consistent with predictions of a source clustering ``signature''.
     This is the first report of such a detection in the far-IR.
     \keywords{cosmology: diffuse radiation, infrared: general} }
   \titlerunning{Background Fluctuations in MIPS 160 $\mu$m Maps}

\maketitle
%
\section{Introduction}
The diffuse cosmic far-IR background emission (CFIB) is believed to be
due to the ensemble emission from galaxies too faint to be resolved;
the spectrum, intensity, and fluctuations of the CFIB across the sky
therefore contain information about the distribution of galaxy
emission in space and time.  Far-IR number counts require that rapid
evolution takes place in IR emitting sources (e.g. Matsuhara et
al. 2000); CFIB fluctuation observations, in conjunction with these
number counts, give more detailed information on the form of the IR
galaxy luminosity function and its evolution (Lagache, Dole, and Puget
\cite{LDP03}).  Current models have the CFIB dominated by emission
from two galaxy populations, non-evolving spirals and evolving
starbursts, with a rapid evolution of the high-luminosity sources (L
\texttt{>} 3 \ensuremath{\times} 10$^{11}$ L$_{\sun}$) between z $= $
0 and 1.

CFIB fluctuations contain information on the clustering of IR emitting
galaxies.  Below, we use the angular power spectrum of intensity
fluctuations (power vs. \textit{k} in inverse angle units) to measure
the structure in these fluctuations.  A Poisson distributed field of
sources of the CFIB would yield a flat angular power spectrum.  A
distribution of IR emitting galaxies equivalent to that observed with
optical galaxy surveys would yield a power spectrum with a log slope
near $-1.1$ (converting the well-known angular correlation function
result w$(\theta) \propto \theta^{-0.7}$ from, e.g., Connelly et
al. 2002, to a power spectrum slope) due to the clustering of the
galaxies, plus a flat component from Poisson intensity fluctuations.
Perrotta et al. (\cite{Perrotta}) used galaxy population models and
number counts across the IR bands to predict the power spectrum in
detail: a low-\textit{k} excess ( \textit{k} $<$ 0.2 arc min$^{-1}$)
above the Poisson component is predicted for the 170 \ensuremath{\mu}m
CFIB power spectrum due to source clustering. 

\begin{table*}
\caption{Reduced MIPS Map Fields}
\label{table:1}
\centering
\renewcommand{\footnoterule}{}  
\begin{tabular}{ccccccccc} 
\hline \hline
Field & RA$^{a}$ &  Dec$^{a}$ & Area &
 Square Area$^{b}$
 & ISM Backgnd &  Tot. Backgnd & Tot. Backgnd\\
         &      &            &             &                         & -Predicted$^{c}$-           & -Predicted$^{c}$-              & -Measured$^{d}$- \\
 & (J2000) & (J2000) & (deg$^{2}$)  & (deg$^{2}$) & (MJy/sr)  &   (MJy/sr)  & (MJy/sr)   \\
\hline
First Look Survey (extragalactic)& 
 17:18:00  & 
 +59:30:00 & 
 4 & 
 4.02 & 
 2.4 & 
 4.4 & 
 6.23 [5.86]$^{e}$\\

SWIRE Lockman Hole& 
10:47:00 & 
58:02:00 & 
 14.4 & 
 8.47 & 
 1.1 & 
 3.6 & 
 4.58 [4.27]$^{e}$ \\
 \hline
\end{tabular}   
\begin{tabular}{l} 
$^{a}$Position of Field Center, \\
$^{b}$Area of square subsection of map used in our analysis.\\
$^{c}$Mean interstellar medium (ISM) or total background estimated at 160 \ensuremath{\mu}m by SPOT  for epoch of observation.\\
$^{d}$Median of reduced map.\\
$^{e}$Values in brackets are for the offset-corrected maps (see section 2.3); 
this process is not intended to determine an improved absolute flux\\
measurement. Note that for the version 10 BCD data, the measured backgrounds 
for the FLS and SLH maps were 7.4 and 5.5 MJy/sr, re-\\
spectively.\\

\end{tabular}   
\end{table*}

Spatial fluctuations in the CFIB were first discovered with ISO at 170
\ensuremath{\mu}m in a relatively small field (0.25 deg$^{2}$; Lagache
\& Puget \cite{LagachePuget00}, Lagache et. al
\cite{Lagacheetal00}). They have also been detected by others in ISO
fields (Matsuhara et al. 2000) and with IRAS data, after re-processing
(Miville-Deschenes, Lagache \& Puget \cite{Deschenes02}). Thus far,
the power spectra of the fluctuations have been consistent with
Poisson distributions of sources, but in the best of these
measurements with ISO, the fields have been too small to accurately
measure and remove the cirrus contribution in order to observe the
predicted clustering (Lagache \& Puget \cite{LagachePuget00}).  The
Multiband Imaging Photometer for Spitzer (MIPS) instrument (Rieke et
al. \cite{Rieke}) has observed much larger fields than Lagache \&
Puget (\cite{LagachePuget00}) with low IR cirrus emission that are
ideal for the study of fluctuations of the CFIB.

In this paper we used data from the MIPS array with sensitivity
centered at 160 \ensuremath{\mu}m and a bandpass of
\ensuremath{\sim}35 \ensuremath{\mu}m, the most sensitive instrument
to date in this wavelength range.  In the verification region of the
extragalactic First-Look Survey (FLS), approximately 15\% of the CFIB
is resolved into sources at 160 \ensuremath{\mu}m (Frayer et. al 2006)
with MIPS, so the remaining 85\% unresolved extragalactic emission
actually dominates source emission. (Note that larger MIPS survey
regions discussed here have shorter integration times and are expected
to resolve somewhat less of the CFIB; e.g. Dole, Lagache \& Puget
2003.)  At 70 \ensuremath{\mu}m, approximately 35\% of the CFIB is
resolved in the extragalactic FLS with MIPS, so studying the
identified sources directly yields somewhat more information at that
wavelength. The characteristics of the sky at 160 \ensuremath{\mu}m
are relatively favorable for the study of the CFIB.  Population models
show that in observations near 160 \ensuremath{\mu}m, the same
distribution of sources that make up the CFIB is also responsible for
the fluctuations in the CFIB ( Dole, Lagache, \& Puget 2003); studying
the fluctuations therefore allows us to learn about the galaxies that
emit the CFIB, even though the majority of these sources are too faint
to study directly.  At 160 \ensuremath{\mu}m the zodiacal emission is
much weaker relative to the unresolved CFIB than at shorter
wavelengths.  The SPOT observations planning tool provided by the
Spitzer Science Center (SSC) predicts that at the time of most of the
MIPS Lockman Hole observations, the zodiacal light intensity at 160
\ensuremath{\mu}m was about 0.9 times the CFIB intensity; at 70
\ensuremath{\mu}m it is 29 times the CFIB intensity. Finally,
contributions from the cosmic mm/microwave background are also small
at 160 \ensuremath{\mu}m compared to measurements in the sub-mm and mm
bands.
 
In order to take advantage of the capabilities of MIPS for CFIB
fluctuation studies, we have undertaken a program to reduce and
analyze the largest low-background fields observed by the
instrument. This paper describes our efforts to
construct and analyze these large 160 \ensuremath{\mu}m Spitzer MIPS
maps using power spectrum analysis. The basic characteristics of the
map observations that we analyzed are given in Table 1.  The aim of
this project is to produce a new measurement of clustering, very
different from those obtained in the optical, to be used to better
understand galaxy and structure formation; we also aim to produce
better power spectra to further constrain galaxy luminosity functions
and evolution models.

\section*{2. Map Observations and Reduction}

\subsection*{2.1 Observations}

As of this writing, the largest contiguous low-cirrus field with good
coverage is the SWIRE Lockman Hole field (SLH).  We also reduce the
First Look Survey (FLS) extragalactic field, the first released, for
comparison. As can be seen in Table I, the SLH field is considerably
lower in interstellar medium (ISM) background (IR cirrus emission; as
known from IRAS and HI maps prior to observations) and larger in area.


The maps were made in scanning observation mode, with the MIPS arrays
performing simple back-and-forth scans. At the end of each scan
(except in a small fraction of the data), the pointing was stepped by
(148\arcsec/276\arcsec) in the cross-scan direction in the SLH/FLS surveys
(nominally just under half the 160 \ensuremath{\mu}m array width ,
\ensuremath{\sim}9.25 pix, for the SLH/ nominally 85\% of the 160
\ensuremath{\mu}m detector width , \ensuremath{\sim}17.25 pix for the
FLS). All SLH scanning observation sequences, or AORs, were performed
twice back-to-back. (The basic unit of planned observation activity
with Spitzer is an AOR, or Astronomical Observing Request. Here we
refer to the series of actions by the observatory, and the associated
data, as the AOR for brevity.) A simple representation of the scan
pattern is given in Fig. 1 for both maps.

\subsection*{2.2 Instrument Behavior - Challenges with Ge detectors}

\begin{table}
\begin{minipage}[t]{\columnwidth}
\caption{SLH Data and Zero Point Constants \protect\footnote{Variance with offsets: 239335. Variance without offsets: 244011. }  }
\label{table_offsets}
\centering
\renewcommand{\footnoterule}{}  
\begin{tabular}{cc}
\hline\hline
AOR Key & Offset(MJy/sr)\\
\hline

{\raggedright {\small 5177088}} & 
{\raggedright {\small -0.0410045}}\\

{\raggedright {\small 5177344}} & 
{\raggedright {\small ~0.0838553}}\\

{\raggedright {\small 5179136}} & 
{\raggedright {\small -0.0582686}}\\

{\raggedright {\small 5179392}} & 
{\raggedright {\small ~0.0984183}}\\

{\raggedright {\small 5179648}} & 
{\raggedright {\small ~0.0494460}}\\

{\raggedright {\small 5179904}} & 
{\raggedright {\small -0.0894248}}\\

{\raggedright {\small 5180160}} & 
{\raggedright {\small ~0.0509532}}\\

{\raggedright {\small 5180416}} & 
{\raggedright {\small -0.0097748}}\\

{\raggedright {\small 5180672}} & 
{\raggedright {\small -0.0789033}}\\

{\raggedright {\small 5180928}} & 
{\raggedright {\small ~0.0552630}}\\

{\raggedright {\small 5181184}} & 
{\raggedright {\small ~0.0140596}}\\

{\raggedright {\small 5181440}} & 
{\raggedright {\small -0.0669587}}\\

{\raggedright {\small 5184768}} & 
{\raggedright {\small -0.0396491}}\\

{\raggedright {\small 5185024}} & 
{\raggedright {\small -0.1431260}}\\

{\raggedright {\small 6592512}} & 
{\raggedright {\small ~0.0085684}}\\

{\raggedright {\small 6592768}} & 
{\raggedright {\small -0.1241370}}\\

{\raggedright {\small 6593536}} & 
{\raggedright {\small ~0.0775429}}\\

{\raggedright {\small 6593792}} & 
{\raggedright {\small -0.0734718}}\\

{\raggedright {\small 6594048}} & 
{\raggedright {\small -0.0432453}}\\

{\raggedright {\small 6594304}} & 
{\raggedright {\small ~0.0090445}}\\

{\raggedright {\small 6595072}} & 
{\raggedright {\small ~0.0809884}}\\

{\raggedright {\small 6595328}} & 
{\raggedright {\small -0.0681358}}\\

{\raggedright {\small 6596096}} & 
{\raggedright {\small -0.0501895}}\\

{\raggedright {\small 6596352}} & 
{\raggedright {\small -0.0108055}}\\

{\raggedright {\small 7770368\protect\footnote{These two AORs are from the validation scans, taken in 
a different epoch from the rest of the data, when the estimated 
zodiacal light contribution was 0.22 MJy/sr lower.}}} & 
{\raggedright {\small ~0.4639960}}\\

{\raggedright {\small 7770624$^b$\label{These two AORs are from the validation scans, taken in 
a different epoch from the rest of the data, when the estimated 
zodiacal light contribution was 0.22 MJy/sr lower.}}} & 
{\raggedright {\small ~0.3312270}}\\

{\raggedright {\small 9628672}} & 
{\raggedright {\small ~0.0402081}}\\

{\raggedright {\small 9628928}} & 
{\raggedright {\small ~0.0327531}}\\

{\raggedright {\small 9629440}} & 
{\raggedright {\small -0.0725303}}\\

{\raggedright {\small 9629952}} & 
{\raggedright {\small ~0.0269623}}\\

{\raggedright {\small 9630208}} & 
{\raggedright {\small -0.0331634}}\\

{\raggedright {\small 9630464}} & 
{\raggedright {\small -0.0134986}}\\

{\raggedright {\small 9630720}} & 
{\raggedright {\small -0.0330788}}\\

{\raggedright {\small 9630976}} & 
{\raggedright {\small -0.0005323}}\\

{\raggedright {\small 9631744}} & 
{\raggedright {\small ~0.0146916}}\\

{\raggedright {\small 9632000}} & 
{\raggedright {\small -0.0804646}}\\

{\raggedright {\small 9633280}} & 
{\raggedright {\small -0.0644805}}\\

{\raggedright {\small 9633536}} & 
{\raggedright {\small ~0.0322532}}\\

{\raggedright {\small 9633792}} & 
{\raggedright {\small -0.0983279}}\\

{\raggedright {\small 9634048}} & 
{\raggedright {\small ~0.1191300}}\\

{\raggedright {\small 9634304}} & 
{\raggedright {\small -0.0092235}}\\

{\raggedright {\small 9634560}} & 
{\raggedright {\small ~0.0412873}}\\

{\raggedright {\small 9634816}} & 
{\raggedright {\small ~0.0012663}}\\

{\raggedright {\small 9635072}} & 
{\raggedright {\small -0.0397363}}\\
\hline
\end{tabular}
\end{minipage}
\end{table}

   \begin{figure}
   \centering
  \includegraphics[width=8.8cm]{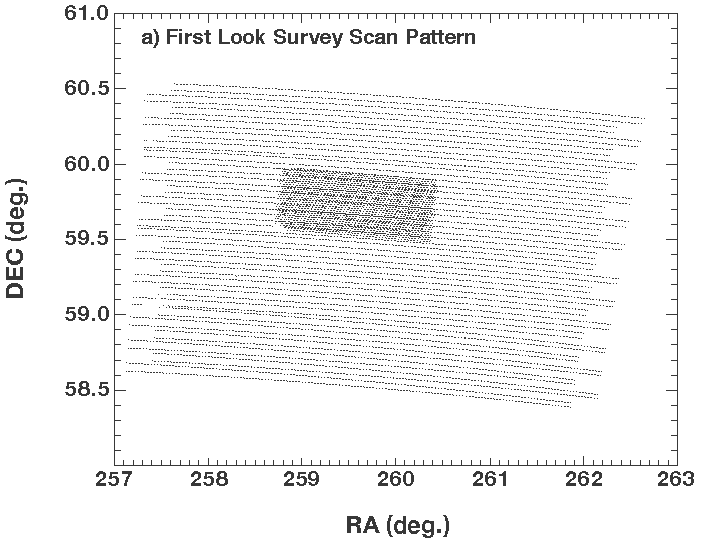}
  \includegraphics[width=8.8cm]{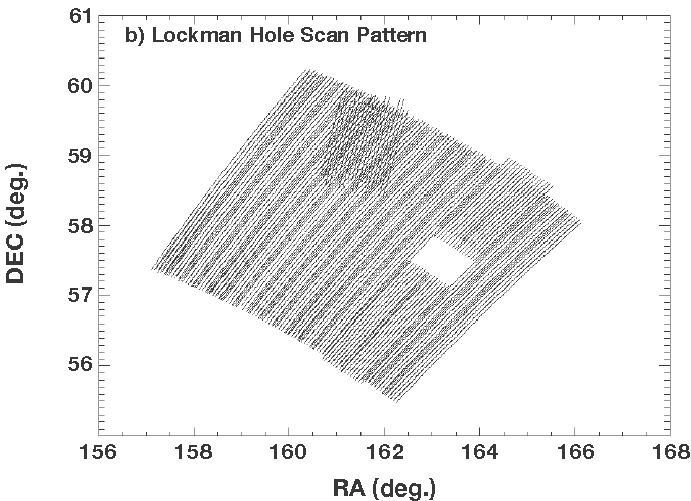}
      \caption{Scan Patterns.  A simple representation of the scan pattern, 
a line connecting each pointing of the camera sequentially, is 
given above for both fields. Part (a) shows the scan pattern for 
the First Look Survey (FLS) Extragalactic Field. Note that all scan paths are close 
to parallel, even those in the verification region (repeated 
observations denoted by denser region at center). Part (b) shows 
the Lockman Hole (SLH) Scan Pattern. Note that all scan paths 
are close to parallel except for those in the validation region 
(repeated observations denoted by dense region of paths at approximately 
45 degrees to nearby paths, at upper middle of figure). 
              }
         \label{figflspointings}
   \end{figure}
%


The MIPS camera 160 \ensuremath{\mu}m measurements are made via a
stressed Ge:Ga detector array.  The response of these detectors is
measured frequently using flashes (``stims'') from a light source
within the camera at a regular period.  Ge detectors are subject to
random and 1/f noise components, including gain drift, and ``memory''
effects, which are extremely difficult to model and correct. (The
so-called memory effect refers to detector responsivity changing as a
function of the history of flux the detector has been exposed to.)  In
practice, drift and memory effects are mostly, but not perfectly,
corrected.  In more typical observations of sources, using scanning or
rapid chopping techniques, the rapid appearance and passing of the
sources in a given detector pixel limits drift and memory effects to
those associated with short time constants.  The stims do a good job
of tracking the detector response on short time scales, and a standard
reduction described in the Spitzer Science Center (SSC) Data Handbook
yields excellent results for both point sources and bright extended
sources.  The MIPS Ge detectors are not optimized for background
observations, however.  Here long time constant effects (``slow
response behavior'') can complicate the reduction and interpretation
of the data.  Below, we describe modifications to the Data Handbook
procedures that we found necessary to obtain good results.  Despite
that fact that the instrument is not optimized for background
observations, we shall demonstrate below that the problems associated
with our data are manageable, and a great deal of information is
available in these maps.

\subsection*{2.3 Basic Map Reduction}

We start with BCD or Basic Calibrated Data from the SSC pipeline
reductions (See Gordon et al. 2005 and MIPs Data Handbook).  We began
our reduction by following the procedures listed in the Data Handbook
for reduction of extended sources, but found that modifications were
required (described below).

Our co-added maps, which have pixels of the nominal camera pixel 
size, give the average of the flux measurements closest to each 
pixel center. No re-sampling of the maps and no distortion corrections 
have been applied at this time because we are interested in structures 
much larger than the camera pixel size. 


   \begin{figure}
   \centering
   
  \includegraphics[width=8.8cm]{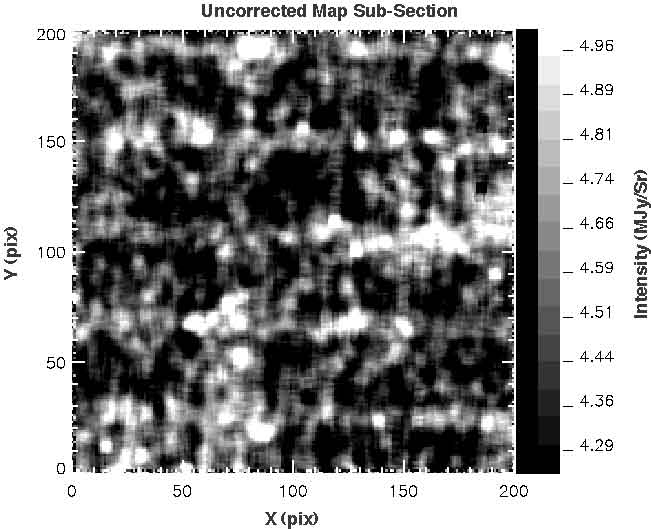}
  \includegraphics[width=8.8cm]{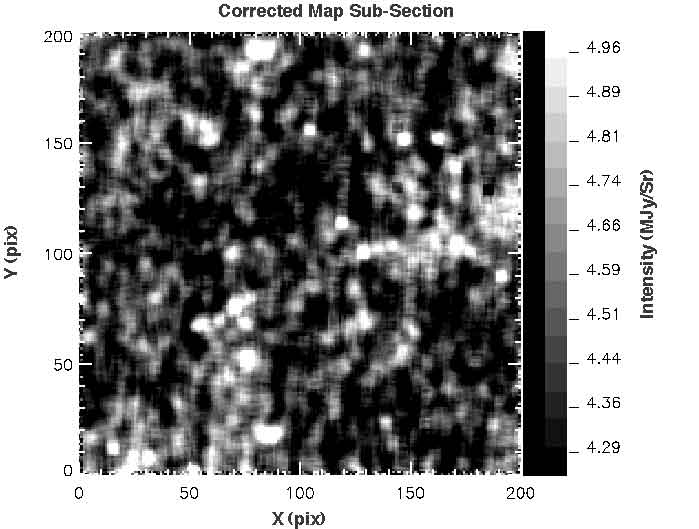}
  \caption{Stim Latent Correction. The two figures show a small
    sub-section of the co-added SLH map.  The top figure shows the
    uncorrected sub-section, which has regular, bright horizontal
    bands (perpendicular to the scan direction) which dominate the
    structure. These bands are due to the stim latent effect. The bottom
    figure shows the corrected sub-section; the bands have been
    completely removed by the correction process. The intensity scale
    (linear, MJy/sr), given at the right side of the figure is the
    same for both images; the spatial scale markings on the bottom and
    left sides of the figure are in units of instrumental pixel
    widths. All data from every field show the same effect.  A 5 pixel
    boxcar smooth has been applied to improve clarity for publication.
  }
         \label{figlatent_cor_vs_none}
   \end{figure}
%

\textbf{Data Selection} This work includes data from pipeline version
11, which includes previously embargoed data \footnote{In our initial
reductions and pre-publication versions of this paper, the data were
processed by the SSC pipeline version 10.  The version 11 data show
improved quality.}.  
Some data covering our selected fields was excluded from our analysis
due to data quality or instrument settings.  In the SLH maps, a
rectangular patch of sky near
(\ensuremath{\alpha},\ensuremath{\delta}) = (163.5\ensuremath{^\circ},
57.6\ensuremath{^\circ}) is covered in AORs 9632512 and 9832256.
These data have high noise and so were not used in our map, leaving a
small rectangular region without data, which we refer to as the
``window''.  These data sets clearly dominated our rms noise map,
described below, with a median rms of 0.33 MJy/sr in these regions vs.
0.23 MJy/sr typical of the rest of the map. (The noise also had
unusually strong structure in the scan direction.) We also excluded
data from PID81, which were taken with different instrument settings
(including stim period) and were not appropriate to combine with the
rest of our data. Table 2 gives a list of AORs which identify the data
used. In the FLS map, all data were used.

   \begin{figure}
   \centering
   
  \includegraphics[width=8.8cm]{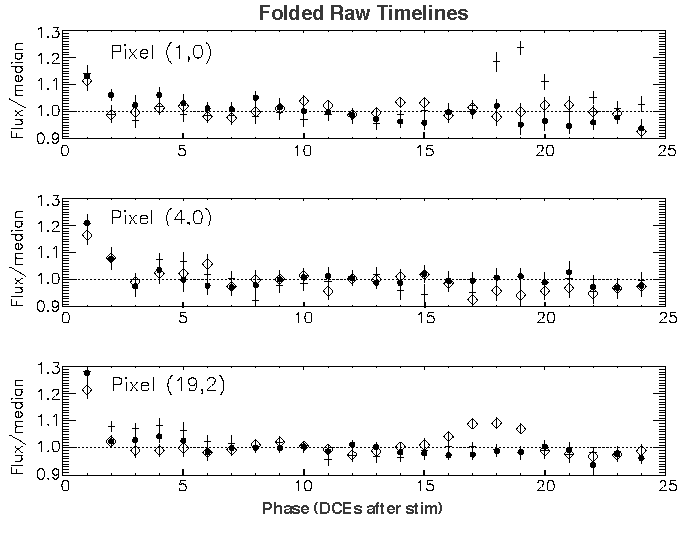}
      \caption{Stim Latents. The data from three different 
detector pixels are shown folded at the stimulator flash period. 
To show possible time-dependence of the phenomenon, the first 
1/3 of the data were represented by a cross, the second third 
by a diamond, and the final third by a filled circle. The time 
bin after the stim flash (which occurs in bin 0 in the figure) 
is almost always high in almost all channels. (These data are 
from AOR 5177088.) The variation within an AOR is small compared
to the other sources of deviation in the data. The response for a given
detector pixel is also consistent between different AORs (with a
well-defined average and some variation for different input sky).
However, each detector pixel has it's own characteristic response
curve, hence our independent correction for each pixel.    }   
         \label{figlatents}
   \end{figure}
%


   \textbf{Stim Correction} In both ``raw'' maps (direct co-adds of
   BCD), obvious parallel bands can be seen perpendicular to the scan
   pattern (See Fig. 2). These correspond to the frames taken just
   after the stim flash response measurements; the effect is referred
   to as a ``stim-flash latent''. The stim flash itself necessarily
   contributes to memory effects in the detectors, as would any
   illumination. These effects depend on the integrated illumination
   history, not just the instantaneous brightness.  Although the flash
   is very bright compared to typical source and background fluxes, it
   is very short in duration, producing small integrated fluence, and
   so causes only a small memory effect.  Unfortunately, the small
   stim latent is significant compared to the faint CFIB.

  Figure 3 shows BCD light curves from the observations of the 
SLH field at 160 \ensuremath{\mu}m (AOR 5177088) folded at the stim 
period. Here, each period of data was divided by the median of 
the data during that period in order to remove variations of 
the sky level during the observation. As can be seen in the figure,
for a given pixel, the behavior is fairly consistent within a
single AOR, but individual pixels can be significantly
different from each other.  Examination of all such response curves
among all AORs shows that the behavior follows normal statistical
variation expected for varying input sky, with a well-defined average
for every time bin for every pixel. We therefore applied a single stim latent
correction to all data, but with a separate correction for each pixel.  
In most pixels, the residual effect is 10\% - 20 \% in the time bin or
DCE after the stim flash, much greater than the error bars from the
variation in sky flux  (3-5\%; see Fig. 3). The correction is simply
the inverse of the folded and normalized timelines.  A comparison of
the sky maps made with and without the stim-flash latent corrections,
shown in Fig. 2, is dramatic. The dominant  structure in the raw map,
the lines made by the stim-flash latents,  has been removed.


\textbf{Illumination correction}
The Spitzer Data Handbook recommends that an illumination 
correction be made for extended source observations.  We therefore 
performed corrections similar to those for CCD flat correction. 
We found the median value for each detector over a large
set of data; this median ``image'' was then normalized, 
and all data in the sample were then divided by the median image. 
The standard deviation of the pixel corrections was typically 7\%. 
We tested making a different illumination correction for each 
scan (i.e. correcting only the data between each change of scan 
direction, as recommended in the Data Handbook), and also using 
the same correction for an entire AOR. If an illumination correction 
were beneficial, we would expect that the rms deviation between 
repeated measurements of the same sky would decrease. In both 
cases, no significant decrease in rms deviation was achieved.

\subsection*{2.4 Zero-Point Correction}

%
   \begin{figure}
   \centering
 \includegraphics[width=8.8cm]{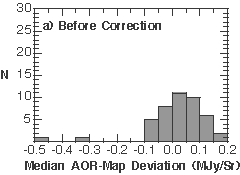}
 \includegraphics[width=8.8cm]{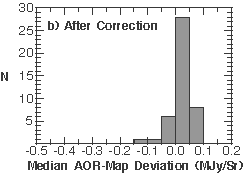}
 \caption{SLH Median AOR Deviation.  In part a), above, the histogram
   of median deviation (median of data $-$ map) for each AOR in the
   SLH map is shown.  The two outliers are 770624 and 770368, the
   validation region scans ($-$0.34, $-$0.48).  In part b), below, on
   the same horizontal scale, the histogram of residual median
   deviations is given (i.e. \textit{after} the optimal set of offsets
   was added to the data).  The added offsets make the distribution
   much more narrow.  The two outlying points are again AORs 770624
   and 770368 ($-$0.08, $-$0.11).  
 }
         \label{deviations}
   \end{figure}
%

In our initial co-added maps, regions associated with a given 
AOR appeared to have discontinuous flux on the borders of regions 
covered by other AORs. On further investigation, we found that 
this was reflected in a systematic discrepancy between measurements 
of the same sky position during different AORs. This can easily 
be seen in the histogram of median AOR data-map deviation. (We 
define the deviation \ensuremath{\delta}$_{i}$ of each measurement \textit{f}$_{i}$ in a 
given AOR at a sky location \textbf{r}(i) to be the difference between 
that measurement and the map at the location of the measurement, 
or \ensuremath{\delta}$_{i}$= \textit{f}$_{i}$(\textbf{r}(i)) - map(\textbf{r}(i)). 
The map value is just the average of all flux measurements \textit{f} at any given 
position, map(\textbf{r}(i)) = \texttt{<}\textit{f}(\textbf{r}(i))\texttt{>}. The median 
deviation for a given AOR is then D = median(\ensuremath{\delta}$_{i}$) for 
all i in the AOR.) The histogram of median AOR deviation is shown 
in Fig. 4 for our selected data. The obvious outliers in the 
SLH data are from observations of the ``validation
region'' of the survey. The validation region data were taken 2003 
December 9, long before the rest of the survey, 2004 May 4-9. 
According to the SSC tool SPOT, the level of zodiacal light is 
estimated to be 0.22 MJy/sr lower during the validation observations. 
However, the deviation in the other data sets taken at the same 
time is apparently instrumental in nature.

We used numerical techniques to find the minimum variance (of 
repeated observations of the same sky pixel) set of additive 
zero point constants to reduce this problem.  The set of minimum 
variance constants 
is given in Table 2 for our SLH map.  We note that this process 
changes the zero-point flux of the map (as noted in Table 1); 
while appropriate for investigations of fluctuations it is not intended for 
correction of the absolute flux zero-point of a map.  

   \begin{figure}
   \centering
   
  \includegraphics[width=8.8cm]{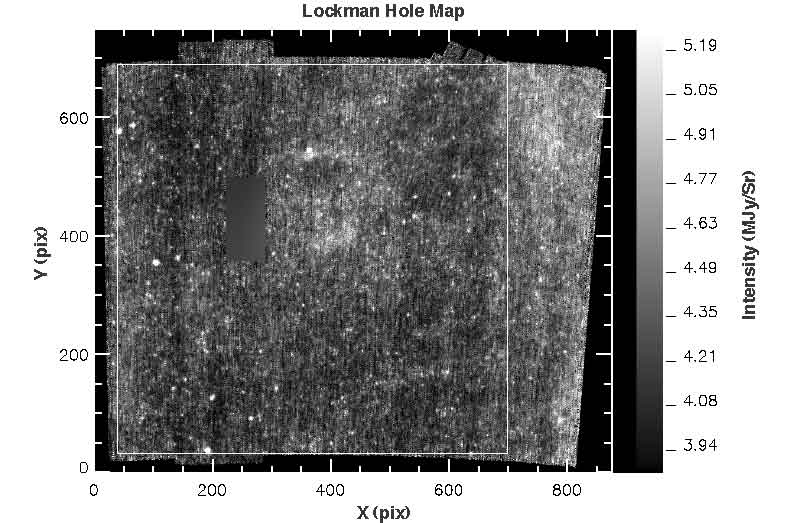}
  \caption{ Full SWIRE Lockman Hole (SLH) 160$\mu$m Map (Scan Direction
    Vertical).  The full SLH map is shown above.  The small
    protrusions at the upper right are the edges of the validation
    region scans, which are roughly 45\ensuremath{^\circ} from the
    main scans. The white rectangle indicates the square sub-region
    used in our analysis.  Because of the strong cirrus emission at the right 
    side of the image, this region was
    omitted from our analysis sub-region. The unnaturally smooth
    rectangle centered near 250, 425 is the ``window'' referred to in
    the text with interpolated data (see text section 3.1).  The x and
    y spatial scales are in pix (15.9"/pix).  The bar at right
    indicates an intensity scale, in MJy/sr.  The image is rotated so
    that the main scans are approximately vertical.  }
         \label{figlhmap}
   \end{figure}
%


\textbf{Zodiacal Light}
No zodiacal light correction was applied (except in the correction 
for a different observation epoch, as described above); we show 
in Section 3, Map Power Spectra, that this produces 
negligible effects on our results.


Figures 5 and 6 show the SLH and FLS maps, respectively, with all the reduction 
steps described thus far.

   \begin{figure}
   \centering
   
  \includegraphics[width=8.8cm]{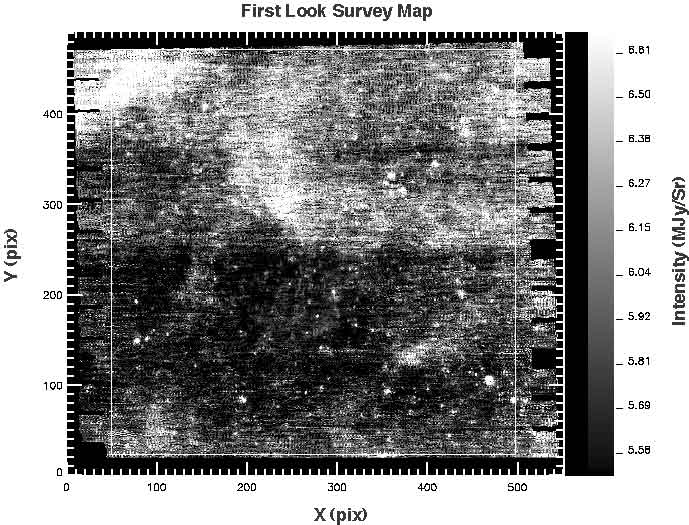}
  \caption{First Look Survey (FLS) Map (Scan Direction
    Horizontal). The full FLS map is shown above. The
    verification region, where additional observations were made, is a
    small region just below center.  The same spatial and intensity
    units are used as in the previous figure. The figure is rotated so
    that the scans are approximately horizontal.  AOR boundary-related
    structure is evident. The overlap of adjacent regions covered by
    different AORs is smaller than in the SLH map, and the average
    number of observations per sky pixel is lower, making good offset
    corrections particularly challenging.  The cirrus flux is also
    higher.  (Because of the poor map quality and high cirrus, this
    map was not used in our final CFIB spectrum or related
    conclusions.  The white rectangle indicates the sub-section used
    in the power spectrum of this region shown in Fig. 10.  See text
    for details.)  }
         \label{figflsmap}
   \end{figure}
%


{\subsection*{2.5 Striping}}

Striping is a common problem in intensity mapping.  When a detector is
moved on a contiguous path, slow changes in detector
response (``drift'') will result in structure in the map
related to the scan pattern which looks like stripes.  Our maps have
such structure.  This can be seen clearly on a computer screen, though
it often shows up poorly in printed reproductions. The scans are all
approximately parallel in the maps (except for small validation
regions), enhancing this structure. In Fig. 7 we show the SLH map
smoothed at a size of 1/2 the instrument width to emphasize this
effect; the figure has been rotated so the scans are vertical, and the
structure is also vertical. The structure in the map looks like large
blocks, the regions covered by each AOR, rather than thin stripes.

Monotonic drift in most detectors in the array, zero-point offsets in
most detectors, and memory effects can contribute to striping.  We
compared the individual detector timelines to the average of all
measurements at the same map points, and determined that there was no
significant linear drift in any pixel timeline. Detector
memory effects are extremely difficult to test for, however, and the
contribution of this effect remains unknown. Detector memory effects
can cause bright extended emission regions to appear ``smeared''
across the map in the scan direction, resembling stripes. When the
instrument finishes scanning a region and points to a new one, the
detector is usually annealed, and its flux history is ``reset''. The
background in the new region will reflect the detector's new history
while in that region, which will be different than in surrounding
regions; hence the discontinuities at the boundaries of the regions
covered in each AOR could be caused by memory effects.

Whatever the cause(s), the residual AOR deviations we showed above
(Fig. 4b) measure the lowest order (most likely dominant) effect in
the timeline data, a constant-per-AOR offset term that causes
stripes. Because no statistically significant linear drift is present
in the timeline data in either the individual channels  or in their
averages during each AOR, we assume the offset-like term dominates the
striping effect.

   \begin{figure}
   \centering
   
  \includegraphics[width=8.8cm]{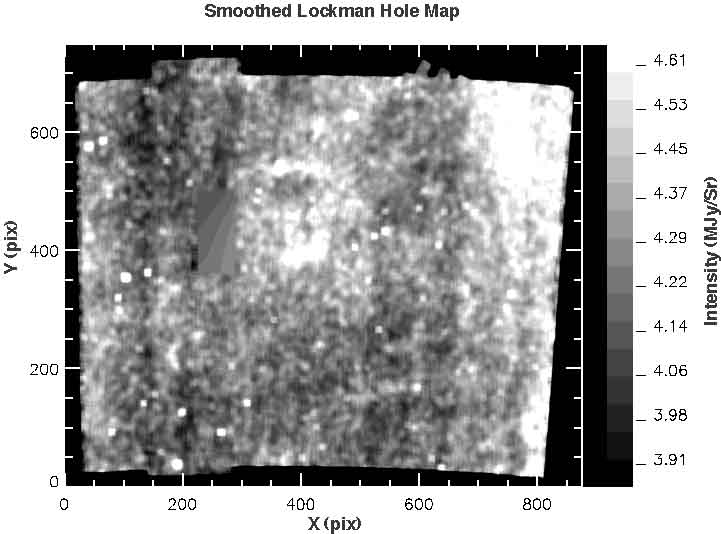}
  \caption{ Full Lockman Hole (SLH) Map Smoothed by 1/2 Detector Width
    (Scan Direction Vertical).  This image of the map has been
    smoothed 
    to enhance structure in the scan direction. Note that in the
    validation region there is significantly less ``smearing'' than in
    the rest of the map due to sampling in different scan directions,
    i.e. cross-linking.  The same spatial and intensity units are used
    as in the previous figure.  }
         \label{figsmoothed_map}
   \end{figure}
%

There is one region where significantly less striping occurs in the
SLH map, in the validation region (see Fig. 5 and Fig. 7), where some
scans were done at large angles to others. (Such scans are said to be
``cross-linked'', described below.) Visual inspection of the maps
shows the worst striping at the extreme right edge of the map in
Fig. 5 and Fig. 7, where several of the brightest, more extended
emission regions are located. Much of this region was not included in
our power spectrum analysis below, however, due to the known (a
priori) high cirrus flux which will tend to dominate the CFIB signal.

\section*{3. Power Spectrum Analysis}


\subsection*{3.1 Power Spectrum Calculations}

Our analysis closely follows that of Lagache et al. (2000). The 
basic steps for analysis of the CFIB are: (1) A power spectrum 
is made from the map. (2) Noise is subtracted from the power spectrum, and 
it is corrected for instrumental response. (3) 
The local foregrounds are then subtracted to yield 
the power spectrum of the CFIB. This section will cover all steps 
of this analysis, and the results will be covered in the following 
section. 

We analyze structure using a simple two-dimensional discrete 
Fourier Transform on square map sub-sections (shown in Fig. 5 
for the SLH map). We report only the average magnitude squared 
of the Fourier components P(\textit{k}) in binned \textit{k} intervals 
(\textit{k} =(k$_{x}$$^{2}$ +k$_{y}$$^{2}$)$^{1/2}$). 
We did not apodize our maps prior to 
power spectrum analysis in order to preserve the information 
in the corners.


 In order to measure the noise power spectrum, two separate maps 
were made from the alternating (i.e. even and odd) measurements 
at each sky location. The even and odd maps were then subtracted 
to make a difference map which was analyzed to determine the 
noise power spectrum. 

\subsection*{3.2 Map Preparations}


\textbf{Missing Data} Certain artifacts of the maps had to be ``repaired'' 
before proceeding to power spectrum analysis. Our SLH field map 
contains an approximately rectangular section, the ``window'', 
0.29\ensuremath{^\circ} \ensuremath{\times} 0.63\ensuremath{^\circ} 
(66 \ensuremath{\times} 143 pixels) for which all AORs have excessively 
high noise (see section 2.3). Because the region includes only 
a small fraction of the map area (0.18 deg$^{2}$, 2.1\% of our square 
map size), the loss of these data should have only a negligible 
effect on our results. We compared several methods for replacement 
of the data in this region: replacement with contiguous sections 
of data taken ``above'' and ``below'' 
the window, from both ``sides'' of the window, and 
finally we replaced the data with a smooth fifth order polynomial 
fit to the data around the window. The power spectrum results 
were insensitive to the choice of replacement method (or interpolation 
order), and we finally adopted the smooth fit.


There are unobserved pixels in the maps; in the FLS about 1.3 
\%, in the SLH 3 \ensuremath{\times} 10$^{-4}$ of the area in our square map 
was unobserved (in addition to the missing rectangle). We replaced 
all small groups of unobserved pixels with a local median of 
non-zero pixel values with a center-to-center distance of \ensuremath{\leq} 
10 pixel widths. These replacements had minimal effect on the 
final power spectrum.

   \begin{figure}
   \centering
   
  \includegraphics[width=8.8cm]{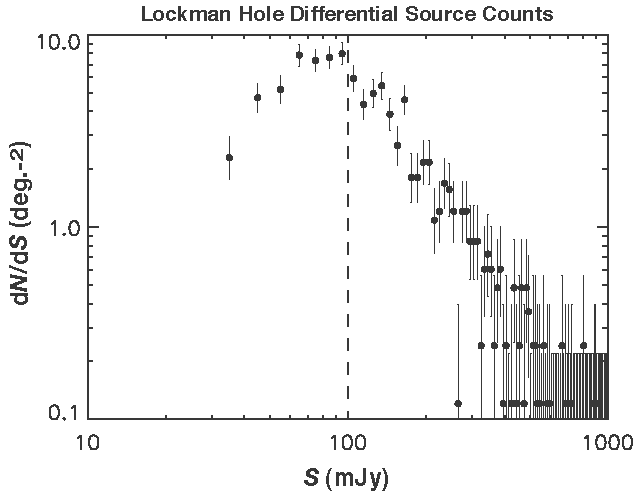}
  \caption{Differential Source Counts for Lockman Hole
        (SLH) Map.  The figure shows differential source counts,
        d\textit{N}/d\textit{S}, where \textit{N} is the number of
        sources per unit area, and \textit{S}
        is source flux (in uniform 10 mJy bins).  We required $S > 4 \sigma$ for all
        sources. Error bars give Poisson  counting 68\% uncertainty.  The
        vertical dashed line indicates the source removal flux cut for
        our CFIB analysis. (See text for additional details).  }
         \label{figsourcehist}
   \end{figure}
%


   \textbf{Source Removal} We removed resolved sources
       from the maps before CFIB power spectrum analysis in order to
       remove the contribution of the brightest sources.  We used a
       simple source-finding algorithm and aperture photometry for
       this goal; for detailed studies of the sources in Spitzer 160
       \ensuremath{\mu}m fields, the reader is referred to, e.g.,
       Frayer et al. (2006) and Dole et al. (2004). Source finding can
       be complicated by the background fluctuations, and following
       the SSC's recommended procedure, we found and measured sources
       in maps made from ``median-filtered data'', effectively
       removing the background before reduction.  Here, for each
       BCD pixel timeline, after applying our stim-latent correction,
       we then subtracted a median timeline from each pixel timeline.
       The median timeline was calculated with a moving window 41 DCE
       in extent.  (The ``FBCD'' median-filtered data from the SSC
       have the same strong stim residuals as the BCD data, and so
       were not useable.)

    We used the SEXTRACTOR program (Bertin \& Arnouts
       \cite{Bertin}) to find source locations (x,y), x and y sizes
       (s$_{x}$, s$_{y}$), and fluxes.  Aperture and color corrections
       were not made.  Sources with isophotes more than 1.5
       \ensuremath{\sigma} (0.53 MJy/sr) above the local background
       were considered to be ``detected'' if they had four or more
       adjacent pixels above the threshold.  The flux distribution of
       sources with measured fluxes greater than 4 times measurement
       uncertainty is shown in figure \ref{figsourcehist}.
       
   \begin{figure}
   \centering
   
  \includegraphics[width=8.8cm]{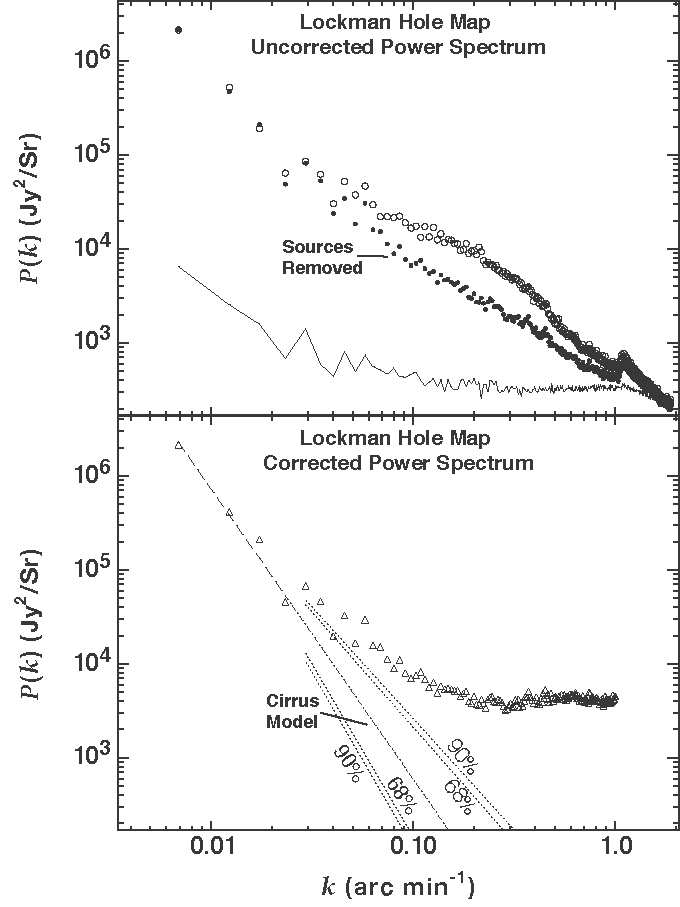}
  \caption{Power Spectrum of Lockman Hole (SLH) Map.  
        In the upper plot, open circles indicate the power spectrum
        (\textit{P(k)}) measured directly from the map; filled circles
        indicate the power spectrum after sources with flux $>$ 100
        mJy were removed.  The continuous line at the bottom of the
    figure gives the measured noise spectrum.  In the lower plot,
    drawn on the same scale for comparison, triangles indicate the
    power spectrum corrected for instrumental response and residual
    zero-point effects.  The power law fit is shown as a dashed line
    in the range of measurements where the fit is made, and its
    extrapolation is denoted by a ``dot-dash'' line (``Cirrus
    Model'').   The dotted lines give the
        two-parameter 68\% (inner set) and 90\% (outer set)
        probability region for the extrapolated cirrus power.}
         \label{figs11bpowspec}
   \end{figure}
%

    Most of our map has 4 (nominal) MIPS passes over
       each sky pixel, however, in a small ``Validation Region''
       another 8 passes were made.  Not only does this region have
       significantly longer integration time, the additional passes
       were made along different directions than the rest of the map
       data, which is expected to reduce systematics (see section
       4.3.2).  In order to assess completeness, we compared source
       detections in this region from maps with and without additional
       validation data.  Sources detected in the ``regular +
       validation data'' maps were missed in the ``regular data only''
       maps with systematically increasing frequency below about 90
       mJy.  We therefore chose 100 mJy as a flux cut value that is
       relatively conservative in terms of completeness, as well as
       convenient for comparison to previous measurements.  For S
       $\geq$ 100 mJy, and requiring measurements at greater than 4
       sigma significance for detection, 37 sources were detected in
       the regular data out of 40 sources detected with the additional
       validation data in the same region.  Within the limitation of
       small number statistics, and assuming the deeper map is 100\%
       complete to 100 mJy, our source lists are greater than 90\%
       complete.  This serves as a rough measure of completeness, but
       note that detection efficiency is expected to vary with local
       background structure.

   A second check on completeness comes from a
       comparison of our source density with those of other
       measurements. For our full map, for S$\geq$ 100 mJy, we measure
       506 sources for 61.1 sources/$deg^2$ ($2.00
       \ensuremath{\times} 10^{5}$ sources ${Sr}^{-1}$).  This number is
       midway between the different measurements in Dole et
       al. (2004), which may be read directly off their figure (see
       top of Fig. 3 in this work; corrections for completeness were
       not made), and is also consistent with Frayer et al. (2006).

   At the location of each source centroid, a
       circular region within diameter d$_{replace}$ was replaced with
       local background values. The resulting one-dimensional source
       sizes s = maximum (s$_{x}$, s$_{y}$), yielded good results as
       d$_{replace}$ in the range of 1.5-3.5 times the instrument
       full width at half maximum, FWHMi (2.375
       pix), except for sources with larger measured sizes, which
       yielded better results when the size was truncated 
       to between 3.5 and 4.5 FWHMi\footnote{Our algorithm
         for determining d$_{replace}$ proceeds as follows: (1) We
         first set d$_{replace} = s$, except where $s <$ 1.5 FWHMi
         where we set d$_{replace} =$ 1.5 FWHMi. (2) For sources with
         flux \texttt{<} 150 mJy and $s > $ 3.5 FWHMi, we truncated by
         setting d$_{replace}$ = 3.5 FWHMi.  (3) For higher flux
         sources with s \texttt{>} 4 FWHMi we used a log truncation,
         d$_{replace}$ = 4 FWHMi + 1.5 ln(1+(s-4 FWHMi)/2) pix. (4)
         All d$_{replace} >$ 4.5 FWHMi were replaced with our maximum
         value, 4.5 FWHMi.}.  In the SLH maps only, two very large and
       bright sources were masked ``by hand''.  For each pixel in the
       replacement region, the median of an annulus of inner and outer
       diameters 3 d$_{replace}$ and 4 d$_{replace}$ replaced the
       pixel value. In the end, the power spectra were insensitive to
       small variations in d$_{replace}$ and to a wide range of values
       of
       the detection isophote threshold.    \\


\textbf{Sub-Sample Selection} The only significant exclusion in the SLH map
was the region at the far right edge in Fig. 5, which was excluded due to high cirrus.  With this exception, we analyzed the largest possible 
square sub-regions of each map (indicated in Figs. 5 and 6).  The SLH
sub-sample area is 8.47 deg$^{2}$. \\

\subsection*{3.3 Features of the Raw Power Spectrum}
      
   \begin{figure}
   \centering
   
  \includegraphics[width=8.8cm]{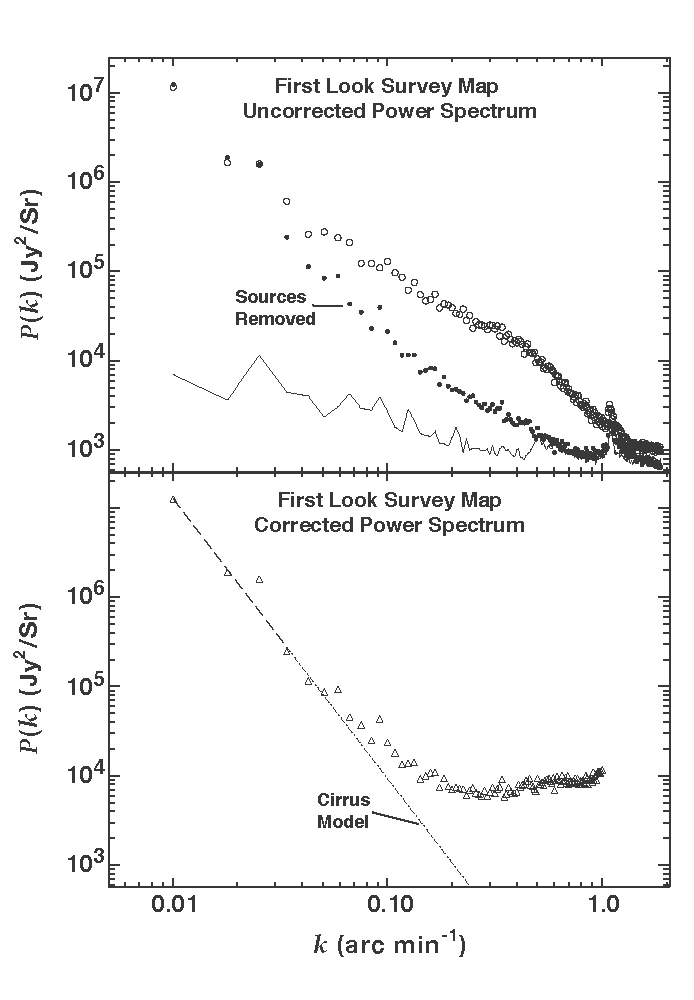}
  \caption{First Look Survey (FLS) Map Power Spectrum.   In the upper
    plot, open circles indicate the power spectrum (\textit{P(k)}) measured directly from
    the map; filled circles indicate the power spectrum after sources
    with flux $>$ 100 mJy were removed.  The continuous line at the
    bottom of the figure gives the measured noise spectrum.  In the
    lower plot, drawn on the same scale for comparison, triangles
    indicate power corrected for instrumental response (PSF).  The
    power law fit is shown as a dashed line in the range of
    measurements where the fit is made, and its extrapolation is
    denoted by a ``dot-dash'' line (``Cirrus Model'').  See text for
    additional details.  Note that the much larger SLH field had
    measurements down to smaller \textit{k} (Fig. 9).  The power law
    extrapolation is not much higher than the data up to $\sim 0.09$ arc
    min$^{-1}$, and so the CFIB cannot be measured easily or
    accurately at low frequency in this field.  }
         \label{figflspowspec}
   \end{figure}
%

\begin{figure}
   \centering
   
  \includegraphics[width=8.8cm]{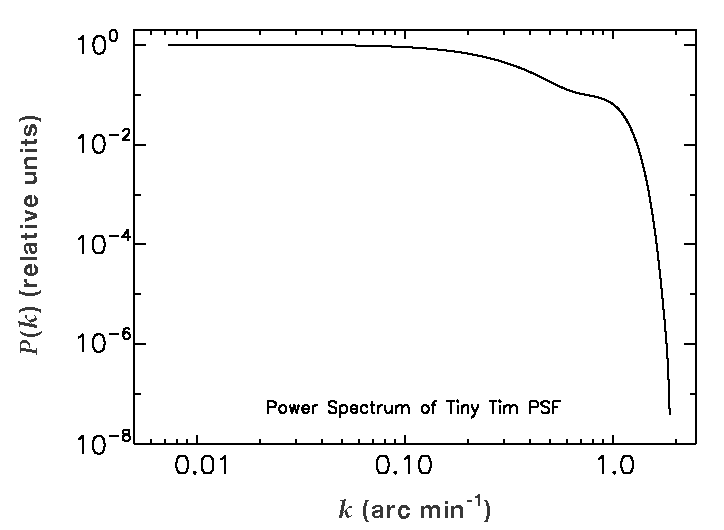}
      \caption{Power Spectrum of PSF.  (PSF generated by the routine Stiny Tim.)        }
         \label{psfpowspec}
   \end{figure}

The power spectra of the SLH and FLS maps are shown in Fig.s
\ref{figs11bpowspec} and \ref{figflspowspec}. The following features
are of interest: The lowest frequency bins look like a simple power
law; in this region IR cirrus emission is known to dominate.  In the
mid-frequencies, there is excess emission above this power law.  This
is the signal due to cosmological sources.  At the high frequency end,
the signal is strongly modulated by the instrumental response
function, the power spectrum of the point spread function (PSF) of the
instrument and telescope. The power spectrum of the PSF provided by
the SSC (simulated by the STINYTIM routine) is shown in
Fig. \ref{psfpowspec}.  At the highest frequencies, the signal becomes
dominated by noise; noise power is within a factor of two of signal by
\textit{k} $\geq$ 0.7 arc min$^{-1}$ for the SLH.

\subsection*{3.4 Systematic Effects}

Below we list likely sources of systematic errors, and demonstrate
that all these errors are small compared to the power spectrum
features we are interested in.

{\subsubsection*{3.4.1 Zodiacal Light}}

The zodiacal light contribution to the power spectrum at 160
\ensuremath{\mu}m is small compared to our CFIB fluctuation signal,
and so we did not attempt to remove any zodiacal light signal from our
maps prior to analysis. At 160 \ensuremath{\mu}m, a simple, planar fit
to the zodiacal background values predicted by SPOT was analyzed to
determine the effect on the power spectrum.  We estimate that zodiacal
light contributes less than 2.5\% of the spectral power in any bin,
with a maximum contribution at low \textit{k}.

{\subsubsection*{3.4.2 Stripe Reduction and Residual Stripe Effects}}

As shown in section 2.4, there are systematic deviations in each AOR
data set that have the character of a zero-point offset.  Such a set
of zero-point offsets might cause false structure because the scan
pattern is rather regular. We tested the effects of such offsets on
the power spectrum by producing a known, ``synthetic sky'', then
simulating observation of this ``synthetic sky'', including the
addition of noise and offsets.  Comparison of the known input
synthetic sky and the resulting maps will then give the systematic
effects due to the zero-point offsets.

We produced our synthetic sky with a very simple Poisson distributed
model CFIB + foreground cirrus. We used an actual cirrus image from
ISSA plates with point sources removed and re-binned to the same pixel
size as in our map. This cirrus foreground image has the desired -3
slope spectrum, but much stronger cirrus than in our field. We scaled
this image in intensity such that it had the same low-k cirrus power
spectrum amplitude as in our SLH map.  We then added a Poisson
distributed synthetic CFIB signal (i.e. with a flat power spectrum),
 such that it produced the same power spectrum amplitude as the flat
high-frequency component of the SLH power spectrum. We then simulated
scanning observations, using the same scanning pattern as in the SLH
observations, adding the measurement noise and offsets as described
for the different simulations below. 

%

   \textbf{Offset Correction Simulation Test:} In our first test, we
   essentially measure the ability of our mapmaking software to
   correct for offsets given the SLH scan pattern and measured noise.
   We started with our synthetic sky simulation and added the same
   noise power as observed in the SLH, and \textit{the actual set of
     median deviations measured in the BCD data}.  Different
   realizations of the simulated observations were achieved by adding
   the set of deviations to the simulated AOR data sets in a
   different, randomly scrambled order (i.e. a given offset was assigned
   to a different region on the map), for each realization.  (We
   generated these different realizations in order to understand
   effects similar to offsets in a very general way.)  We reduced the
   simulated observation data sets in the same way as our real maps,
   fitting for zero-point offset corrections. At this point, most of
   the effects of offsets should have been corrected, and we expect
   little effect on the power spectrum.

Despite realistic added noise, our offset correction routine worked
well, yielding small residual deviations. The power spectrum showed
only a very small effect due to uncorrected offsets ($< 15\%$ for $k<
1$ arc min$^{-1}$), negligible compared to the uncertainty due to the
fit errors at low \textit{k}.  

\begin{figure}
   \centering
   
  \includegraphics[width=8.8cm]{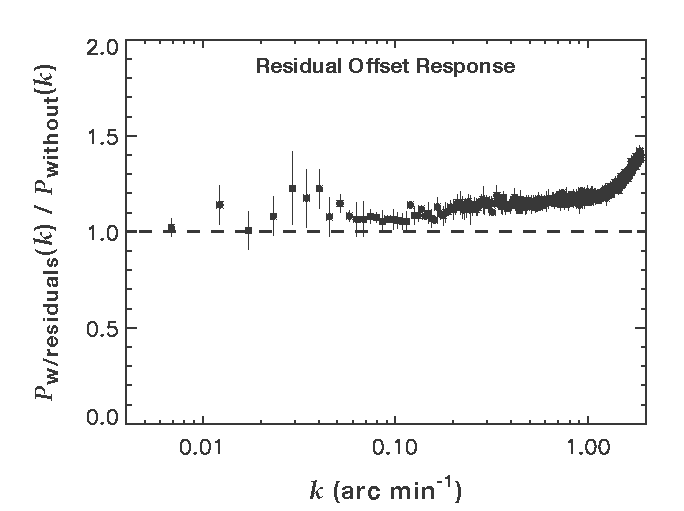}
  \caption{Effect of Residual Offsets on the Power Spectrum.  The
    figure shows the relative effect of the observed residual offsets
    on the power spectrum.  See Section 3.4.2 for details.  }
         \label{residspecfunc}
   \end{figure}

\textbf{Residual Offset (Stripe Effects) Simulation and Correction:}
Here we observe the effects of \textit{uncorrected}, residual offsets
on the power spectrum, to match the offsets seen in Fig. 4b.  Starting
again from synthetic sky maps, we simulated observing on the SLH scan
pattern, added the same random noise power as observed in the SLH, and
\textit{the same set of residual median deviations measured in our
  final SLH map} (see Fig. 4b). Analogous to the procedure above,
different realizations of the simulated observations were achieved by
adding the set of residual deviations to the simulated AOR data sets
in a different, randomly scrambled order, for each realization.  (We
generated these different realizations in order to understand effects
similar to the measured residual deviations in a very general way.) We
reduced these simulations in the same way as our real maps, but of
course, without correcting the residual deviations.  The effect on the
power spectrum is $< 25\%$ for $k< 1$ arc min$^{-1}$ (see Fig.
\ref{residspecfunc}).  These errors are small compared to
uncertainties in the low-\textit{k} power law fit which dominate the
low- to mid-\textit{k} CFIB measurement, discussed in the next
section.  Below, we make use of the function in
Fig. \ref{residspecfunc} to make a correction for the effect of
residual deviations on the power spectrum, the inverse of the function
in the figure.

%

\subsection*{3.5 CFIB Analysis}

The CFIB analysis requires correction of the raw map power spectrum 
for potential effects due to offsets and removal of foregrounds. \\

\textbf{Instrumental Response Correction}
The sky map may be described as the convolution of the real sky 
and an instrumental response function plus noise. The power spectrum 
of the sky may therefore be derived from the instrumental power 
spectrum minus the noise power spectrum, divided by the response 
function of the instrument.  In the analysis below, we assume that the 
PSF dominates the instrumental response function, and approximate 
our instrumental response function by the power spectrum of the 
PSF given in Fig. \ref{psfpowspec}.

   \begin{figure}
   \centering
   \includegraphics[width=8.8cm]{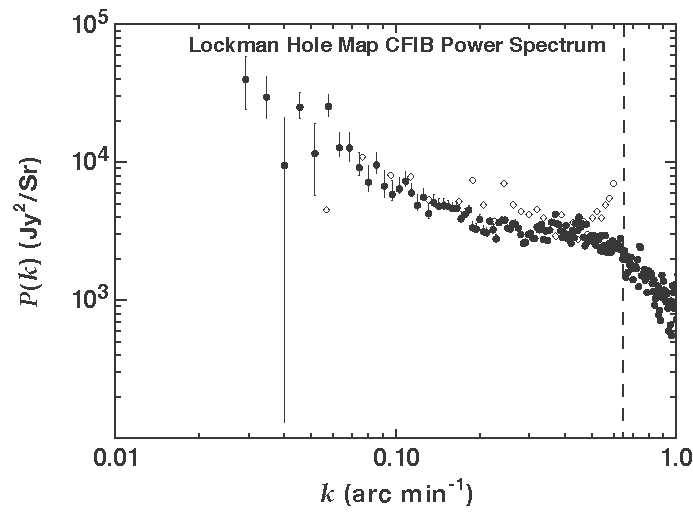}

   \caption{CFIB Spectrum from the SLH (Lockman Hole) Map.  The figure
     shows our noise-subtracted, instrumental response, and residual
     deviation-corrected CFIB spectrum of the SLH subsample (filled
     circles).  The low-frequency power appears to fall rapidly to
     \ensuremath{\sim}0.1 arc min$^{-1}$; there is a comparatively
     flat region \ensuremath{\sim} 0.2 - 0.4 arc min$^{-1}$.  The
     error bars give the 68\% probability intervals for uncertainty due to 
      the cirrus power law
     subtraction and residual deviation correction.  The cirrus
     subtraction uncertainty dominates at low frequencies.  The
     diamonds reproduce the results given in Lagache \& Puget
     (\cite{LagachePuget00}). For \textit{k} $>$ 0.7 arc min$^{-1}$
     (vertical dashed line) the noise becomes within a factor of 2 of the
     signal, and the PSF correction is greater than a factor of 10.  }
         \label{figscfibspec}
   \end{figure}
%


\textbf{Residual Offset Correction to the Power Spectrum}
We also applied a correction for our residual offsets (measured
residual deviations) described in Section 3.4.2.  To correct for these
residual offsets, we divided out the effect of these offsets in our
simulations, the function given in Fig. \ref{residspecfunc}.

\textbf{Foreground Cirrus Subtraction}

Previous measurements of the sky on scales from arc seconds to much
larger than our maps have shown that cirrus structure has a power law
shape with a log slope very close to --3 (e.g., Wright \cite{Wright},
Herbstmeier et al. \cite{Herbstmeir}, Gautier et al. \cite{Gautier},
Kogut et al. \cite{Kogut}, Abergel et al. \cite{Abergel}, Falgarone et
al. \cite{Falgarone}). This very steep slope means that this structure
must dominate at the lowest frequencies (as shown in Fig.
\ref{figs11bpowspec}).  Following Lagache et al.
(\cite{Lagacheetal00}), we subtract a power law fit of the
low-frequency structure from our power spectrum in order to remove the
cirrus contribution.  We fit a power law function to the lowest four
bins of \textit{k} in order to get a good fit of the cirrus structure
in the range of \textit{k} where it dominates.  Unlike Lagache et
al. (2000) we do not assume a power law slope, but fit both amplitude
and slope. Given the extensive evidence for power law cirrus
structure, we decided to empirically determine our fit errors from the
deviation in the log of our data from the even-weighted power law
fit. The dotted lines shown in the bottom plot of
    Fig. \ref{figs11bpowspec} reflect our mapping of 2-d $\chi^{2}$
    space 68\% and 90\% probability contours.  These fit
    errors dominate the uncertainty in our CFIB measurements at
    low-\textit{k}. The resulting CFIB fluctuation power spectrum (the
    result of subtraction of the cirrus fit from the corrected power
    spectrum) is shown for the SLH field in Fig. \ref{figscfibspec}.
    (The upper and lower error bars in the figure are the sum in
    quadrature of the 68\% upper and lower fit uncertainties and the 1
    $\sigma$ errors in the residual offset correction.)

\section*{4. Results}

\subsection*{
4.1 The CFIB Fluctuations Measurement}

{\subsubsection*{4.1.1 The SLH CFIB Fluctuation Spectrum}}

The observed CFIB power spectrum is described in gross terms as
showing high power at low \textit{k}, decaying rapidly to
\ensuremath{\sim}0.1 arc min$^{-1}$, with a relatively flat region
\ensuremath{\sim}0.2 - 0.4 arc min$^{-1}$.  If the sources of the
CFIB were distributed at random in space, a flat power spectrum would
be expected. What is observed is clearly different.  This excess CFIB
power at low \textit{k} has been identified as the signature of
clustering of CFIB sources, discussed in the following section. The
error bars on the lowest few \textit{k} values are large due to the
finite uncertainty in the subtracted power law fit. However, the
remaining points have quite reasonable uncertainties, and we
concentrate on these in our discussion.  At the large \textit{k}
values, the noise becomes significant and the instrumental correction
becomes very large (see Fig. \ref{figs11bpowspec}; the figure is cut
off at \textit{k} = 1.0 arc min$^{-1}$. )

{\subsubsection*{4.1.2 Comparison with Predictions}}

\begin{figure}
   \centering
   
  \includegraphics[width=8.8cm]{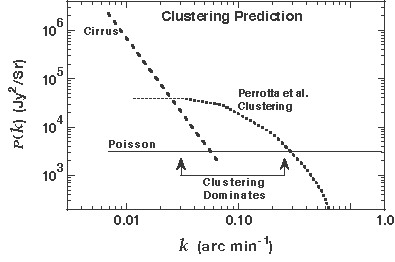}
  \caption{Power Spectrum Clustering Prediction. The
        power spectrum figure above shows the Perrotta et al. (2003)
        clustering prediction (for sources with fluxes $<$ 135 mJy;
        short dash line) taken from Dole, Lagache, \& Puget (2003). We
        also show our Poisson level from the SLH map power spectrum
        (the average power measured between \textit{k} $=$ 0.2 and 0.5
        arc min$^{-1}$, 3203 MJy/Sr; solid line).  The cirrus
        foreground structure fit from the SLH map is also shown
        (longer dashed line).  Following Dole, Lagache, \& Puget
        (2003) a clustering ``signature'' should be detectable between
        roughly \textit{k} $=$ 0.03 and 0.3 arc min$^{-1}$. We see
        an excess in our SLH map power spectrum in about the same
        range, \textit{k} \ensuremath{\sim} 0.03 to 0.15 arc
        min$^{-1}$ (Fig. \ref{figscfibspec}).  }

         \label{figperrotta}
   \end{figure}

Perrotta et al. (2003) showed the effects of predicted clustering on
background power spectra, using a spiral+starburst population, as did
Lagache, Dole \& Puget (\cite{LDP03}).  Lagache, Dole \& Puget
(\cite{LDP03}) constructed simulations of Spitzer shallow survey
observations, but distributing their galaxy populations at random
(without clustering) for comparison.  The simulations used a galaxy
distribution (``normal'' spirals + evolving starburst galaxies) and
evolution constructed to be consistent with IR - mm source counts,
with sources distributed at random, and interstellar foreground
cirrus. By comparing their simulation of cirrus plus unclustered
(Poisson distributed) galaxies to the clustering component predicted
by Perrotta et al. (2003), they were able to predict that the
clustering component would be visible as a bump above the cirrus and
Poisson components in the range of \textit{k} $=$ 0.04 -- 0.2 arc
min$^{-1}$.  In Fig. \ref{figperrotta} (after Dole, Lagache, \& Puget,
2003, Fig. 11) we show the theoretical clustering component
prediction, along with power spectrum components from our SLH map, the
power law fit to the cirrus, and the flat or ``Poisson'' component (the
average flux density measured from 0.2 arc min$^{-1}$ to 0.5 arc
min$^{-1}$, 3203 MJy/Sr).  (Note that Perrotta et al. 2003 remove 
sources down to 135 mJy from their analysis, not 100 mJy as we do here.) 
Looking at Fig. \ref{figperrotta} and
following Dole, Lagache, \& Puget (2003), a clustering ``signature''
should be detectable roughly between \ensuremath{\sim} 0.03 and 0.3 arc
min$^{-1}$, where the clustering power spectrum component is greater
than the cirrus and ``Poisson'' components.

%

In the grossest sense, these predictions do agree with our CFIB
spectrum, in that there is a substantial low-to-mid \textit{k} excess
above a relatively flat high- \textit{k} ``Poisson'' region, the
clustering ``signature'', clearly inconsistent with a Poisson
distribution of sources.  In detail, however, Perrotta et al. (2003)
predicts that the power spectrum excess is almost flat for \textit{k}
in the range of 0.01 to 0.1 arc min$^{-1}$, which is not consistent
with our observations.  Our points with small error bars near
\textit{k} $=$ 0.045 and 0.06 arc min$^{-1}$ require a drop of greater
than a factor of 3 in power between these bins and 0.1 arc min$^{-1}$.
 
{\subsubsection*{4.1.3 Impact of Systematic Effects}}

These results are from a relatively simple analysis which explicitly
assumes that the SSC PSF simulation gives the correct instrumental
response function. However, it is extremely unlikely that an
instrumental feature would produce our excess low-\textit{k}
(large-scale) power.  Large and medium scale power in the instrumental
response function can only come from changes in the instrument over
times comparable to observation of the whole map; the camera PSF has
no structure on large scales (see Fig. {psfpowspec}). The map was sampled rather
uniformly on the large scales, with essentially the same scanning
overlap, redundancy, scan speed, etc. across most of the map.  There
is no strong feature in the power spectrum (except for the
high-\textit{k} PSF signature) common to any set of maps we have
examined that would suggest a significant uncorrected instrumental
artifact.  The ISO instrument (Lagache \& Puget \cite{LagachePuget00})
also had no features in the instrumental response in the low- to
medium-\textit{k} range, and in general, such features would not be
expected. We note that the excess clearly extends into the
mid-\textit{k} region (the power is more than a factor of two above
the flat region all the way up to \textit{k} $=$ 0.08 arc min$^{-1}$).
The map contains many samples of the associated size for mid-
\textit{k} values, therefore the measured behavior cannot be a
statistical anomaly.  We also note that spurious artifacts generally
occur at much smaller spatial scales.


In section 3.4.2 we addressed the possibility that systematics 
with the character of an instrumental zero-point offset could 
cause erroneous structure in our power spectra. The result of 
our simulations showed that after we corrected for offsets, this 
was a small  effect.  (Compare the magnitude of the excess in 
Fig. \ref{figscfibspec} to the magnitude of the corrections in 
Fig. \ref{residspecfunc}.) Further, we applied a correction to the spectrum 
for the residual uncorrected offsets (stripes). The consistency of our simulations 
makes us confident in this correction. In the end, however, the 
erroneous structure and correction had little effect on the results.  Table 3 
lists the power law fits to the SLH power spectrum with and without 
offset and residual spectral corrections, and they are essentially 
indistinguishable. Because 
the dominant error in the CFIB spectrum is the power law subtraction 
uncertainty at the low to mid-\textit{k} points of interest, and because the 
power law fit did not change, neither the effect of the offsets nor the 
corrections significantly changed the results in the low to mid-k points of 
interest.  The excess at low to mid-\textit{k} is simply larger than any
of these effects or corrections.  Our main result of excess power at  low 
to mid-\textit{k} is therefore robust against 
such systematic effects, within our errors. \\
 
%
\begin{table}
\begin{minipage}[t]{\columnwidth}
\caption{Low-\textit{k} Power Law Slopes}
\label{table_powspecresults}
\centering
\renewcommand{\footnoterule}{}  
\begin{tabular}{cc}
\hline \hline
{\centering \textbf{Map and Sample}} & 
{\raggedright \textbf{Cirrus Power Law Slope}\footnote{Power law fit to the first four points of the power spectrum. }}\\
\hline
{\centering SLH,corrected} & 
{\raggedright --3.11 \ensuremath{\pm} 0.30}\\

{\centering SLH,uncorrected} & 
{\raggedright --3.09 \ensuremath{\pm} 0.29}\\

{\centering FLS} & 
{\raggedright --3.12 \ensuremath{\pm} 0.55}\\

\hline
\end{tabular}
\end{minipage}
\end{table}
%

{\subsubsection*{4.1.4 Dispersion in the SLH Results}}

The SLH sub-sample was selected in order to avoid bright cirrus
regions and map edges. We reduced a few different sub-sample regions
of similar size within the full map (Fig. 5) which included varying
amounts of bright cirrus.  As the subsample included more bright
cirrus, we found larger uncertainties in the power law fits and
associated larger CFIB power spectrum errors.  All cirrus-subtracted
CFIB power spectra are consistent within errors, however.

We note that the two SLH validation AORs yielded the largest deviation
even after zero-point correction (Fig. 4). Removal of these two AORs
from the map yielded results consistent with those reported above.
Because we did not find otherwise obviously anomalous behavior in the
region covered by these data in the difference maps, however, we
judged that elimination of these data may not be justified.  We
therefore did not eliminate these data.

{\subsubsection*{4.1.5 First Look Survey Results}}

The FLS map power spectrum has no obvious turn-off from the cirrus
power law until about 0.09 arc min$^{-1}$, whereas in the SLH power
spectrum, the turnoff was at \ensuremath{\sim} 0.03 arc min$^{-1}$.
The difference is due to the greater strength of the cirrus emission
in the FLS field.  The cirrus power is more than 10 times
higher in the FLS at 10$^{-2}$ arc min$^{-1}$ than in the SLH. \\

The power law fit was again made from the first four bins. However, 
the field is smaller, so the first four bins are at higher frequency 
than in the SLH. In this field, the power law fit has larger 
uncertainty than in the SLH.  Because of this larger uncertainty, 
the signal in the low to mid-\textit{k} range of interest has large 
error bars, and the resulting CFIB measurement is not interesting. 
This result is not unexpected, however: the combination of strong 
cirrus in the field (requiring very small fit error for CFIB 
measurement), and sampling the cirrus at a higher \textit{k} (because 
the field is smaller) where the CFIB can interfere makes for 
an \textit{a priori} difficult measurement. \\

%
\begin{table*}
\caption{Power Spectra Compared}
\label{table_powspec_compared}
\centering
\renewcommand{\footnoterule}{}  
\begin{tabular}{cc c|c cc cc c}

\hline \hline
{ \  }    &
\textbf{ This Work} &
{ \  }   &
 { \  }   &
{ \  }   &
{ \  }   &
\textbf{ Other Work } &
{ \  }   &
{ \  }    \\
\hline
{\centering \textbf{ S$_{cut}$ }  } & 
{\raggedright \textbf{ {P$_{0.2-0.5}$}$^{a}$}   } &
{\raggedright \textbf{ {P$_{0.05-0.5}$}$^{b}$} }  &
{\raggedright \textbf{Measurement}  } &
{\raggedright \textbf{  {P$_{Poisson}$}$^{c}$}  }  &
{\raggedright \textbf{Measurement}  } &
{\raggedright \textbf{  {P$_{Poisson}$}$^{c}$}  }  &
{\raggedright \textbf{Measurement}  } &
{\raggedright \textbf{  {P$_{Poisson}$}$^{c}$}  }    \\

{\centering \textbf{(mJy)} }  & 
{   \raggedright \textbf{(Jy$^{2}$/sr)}   }  &
{   \raggedright \textbf{(Jy$^{2}$/sr)}   }&
 { \  }  &
{   \raggedright \textbf{(Jy$^{2}$/sr)}    }&
{ \  }   &
{   \raggedright \textbf{(Jy$^{2}$/sr)}    }&
{ \  }    &
{   \raggedright \textbf{(Jy$^{2}$/sr)}    }    \\

\hline
{\centering 250} &  3920 &  5341  & 
ISO Lockman$^{d}$ & 12000{\raggedright  \ensuremath{\pm} 2000} & &  &  & \\

{\centering 100} &  3203 &  4472  &  
ISO Marano$^{e}$ & 7400{\raggedright} & ISO ELAIS N1$^{f}$ &  5000 & ISO ELAIS N2$^{f}$  & 5000\\

 \hline
\end{tabular}
\begin{tabular}{l} 
$^{a}$Average, even weight per bin, over the relatively flat region of
our spectrum, \textit{k} $=$ 0.2-0.5 arc min$^{-1}$.  \\
$^{b}$Average, even weight per bin, \textit{k} $=$ over 0.05-0.5 arc min$^{-1}$.\\
$^{c}$Result of Poisson sources + Cirrus Fit.\\
$^{d}$Matsuhara et al. (\cite{Matsuhara})\\
$^{e}$Lagache \& Puget (\cite{LagachePuget00})\\
$^{f}$Lagache et al.(\cite{Lagacheetal00})\\

\end{tabular}   

\end{table*}

%

\subsection*{4.2 Comparison with Previous Results}

\textbf{Background Level} The SSC provides a tool to separately
predict the local, ISM, and extragalactic background components (part
of the SPOT software) incorporating HI data and IR measurements
(essentially the Schlegel, Finkbeiner \& Davis (\cite{Schlegel})
results), and knowledge of the MIPS instrument. Table 1 shows that in
all cases, the median total background is much higher than the
prediction. It turns out that there is a known, but undocumented,
constant zero-point flux offset of about 1 MJy/Sr in the MIPS 160
camera, in all modes, essentially an uncorrected dark signal
(Noriega-Crespo \cite{Noriega-Crespo}).  Subtracting the 1 MJy/Sr
offset would greatly improve the agreement between the SPOT
predictions and the measurements.  The SPOT predictions do agree
roughly with measurements from other instruments in this wavelength
regime (e.g. ISO measurements in the FIRBACK fields; Lagache \& Dole
2001), however the predictions are not intended to be precise and in
particular suffer the limitation of not including CFIB fluctuations.


\textbf{CFIB Fluctuations} The Lagache et
    al. (\cite{Lagacheetal00}) ISO CFIB fluctuation spectrum results
    for the ELAIS N2 field at 170 $\mu$m (diamond symbols), with the
    same 100 mJy source removal, are plotted along with our results in
    Fig. \ref{figscfibspec}. (The values were taken directly from
    their Fig. 3.)  The Lagache et al.  (\cite{Lagacheetal00}) values
    are systematically higher ($\sim$ 35\% higher in the
    mid-\textit{k} region), but given the bin-to-bin scatter,
    differences in instruments, and realistic measurement and
    calibration uncertainties\footnote{Lagache \& Puget
      (\cite{LagachePuget00}) did not include a cirrus-subtracted and
      instrument-corrected CFIB spectrum for direct comparison, hence
      our use of the Lagache et al. (\cite{Lagacheetal00}) CFIB
      spectrum in Fig. \ref{figscfibspec} .  We note that essentially
      the same reduction is used in both papers, however, different
      PSF spectra were used for instrumental correction.  The Lagache
      \& Puget (\cite{LagachePuget00}) PSF derived from observations
      of Saturn would yield significantly higher values in the first
      three lowest \textit{k} values, but the rest of the values would
      be everywhere less than 0.3 dex from those shown.}, the
    agreement is as good as can be expected. We note that the Lagache
    et al. (\cite{Lagacheetal00}) values are consistent with the same
    systematic rise toward low-\textit{k} as is seen in our
    measurements.

Most other works do not give a cirrus-subtracted CFIB
    spectrum; they only report an effective average power spectrum
    value, so comparisons are less straightforward.  All the other
    authors assumed the CFIB power spectrum to be flat, which we have
    shown is incorrect. (The average power spectrum value is then
    dependent on the range of \textit{k} over which the average is
    taken, and even possibly on binning.)  In Table 4, we give average
    values from our SLH map spectrum, as well as other published
    measurements.  Our ``Poisson level'' values, which we putatively
    identify with the relatively flat region \textit{k} = 0.2-0.5 arc
    min$^{-1}$, are much lower than any of the other values in the
    table.  We also report, however, an average power over a larger
    range in \textit{k}, 0.05-0.5 arc min$^{-1}$.  This range in
    \textit{k} is closer to that used in the fits of the other
    authors, and gives closer results. The disagreement with Matsuhara
    et al. (\cite{Matsuhara}) in the same region of the sky, however,
    is large.  Such a large difference is likely due to different flux
    calibrations. For the measurements with source removal down to 100
    mJy, the agreement is better; our 0.05-0.5 arc min$^{-1}$ value is
    only 11\% below the values given by Lagache et. al
    (\cite{Lagacheetal00}) for the ELAIS fields, though $\sim$ 30\%
    below the value for the Marano field\footnote{As
          explained in the previous footnote, the disagreement between
          the ELAIS and Marano values is due, at least in part, to
          different corrections in these works.} (Lagache \& Puget
    \cite{LagachePuget00}).  None of these works, however, claims to
    have addressed the background flux calibration uncertainties of
    their measurements. In our case, the MIPS flux calibration
    continues to evolve, and is a likely source of disagreement with
    other instruments.
\\


\section*{5. Discussion}

\subsection*{5.1 Measurement of Clustering}

As noted in the introduction, one may convert structure found in optical
correlation function measurements to an intensity fluctuation angular power
spectrum.  The result is a power law excess at low-\textit{k}
due to clustering, falling to a flat Poisson component at some higher
\textit{k}.  The Perrotta et al. (2003) prediction, explicitly for the
170 $\mu$m CFIB fluctuation power spectrum, also predicts a
low-\textit{k} excess due to clustering.  In gross detail, this is
what is observed, and is therefore no surprise.  This detection of
clustering structure is also robust because the systematics have been
explicitly controlled: first, our simulations show that there can be
no significant low-frequency distortions in the power spectrum due to
offset-like effects, and second, because PSF and other instrumental
corrections are small at low-\textit{k}.

Now let us consider the shape of the CFIB power spectrum in more
detail.  Though our results are similar to the power spectrum
predicted by Perrotta et al. (2003), \textit{in detail} our results
are clearly different.  The steeply descending low-\textit{k} spectrum
we observe would be very difficult to fit to the nearly flat excess
predicted by Perrotta et al. (2003) in this region. This predicted
flatness comes from the details of the population model used for the
prediction: At low \textit{k} the Perrotta et al. (2003) power spectrum is
dominated by the contribution from starburst galaxies, which is a
nearly flat for \textit{k} $<$ 0.1 arc min$^{-1}$.  This prediction has
spiral galaxies, contrarily, having a power spectrum contribution that
falls steeply with \textit{k}, precisely what is required to reproduce
our sharply falling spectrum at low \textit{k}.  If the power spectrum
contributions of the two galaxy types were correct in shape but not in
strength, then the contribution of spirals would need to be much
larger than that in Perrotta et al. (2003), about a factor of 7
larger.  (We made a crude fit of the Perrotta et al. 2003 spiral
contribution to our low-\textit{k} data, taking the power spectrum
values from their Fig.  7, and determined that the spiral contribution
would need to be 7 \ensuremath{\pm} 2 times as great to be consistent
with our data. We fit in the range of \textit{k} $=$ 0.07 to 0.1 arc
min$^{-1}$.)

The Perrotta et al. (2003) paper has some limitations for comparison
in detail to the results here. Their analysis specifies sources
removed down to 135 mJy, slightly different than our cut at 100 mJy.
The model galaxy distribution used in the prediction is also not
consistent with the most recent Spitzer results.  For example, their
model predicts 3.9 \ensuremath{\times} 10$^{5}$ sources per square
degree at 50 mJy at 175 \ensuremath{\mu}m, which is in poor agreement
with the MIPS 160 $\mu$m measurement of (8.5 \ensuremath{\pm} 1.4)
\ensuremath{\times} 10$^{5}$ sources per square degree (Dole et
al. 2004; the uncertainty is the standard deviation of the two values
measured on different fields).  The log slope of N($>$S) in this
region is --2.3 in Perrotta et al. (2003); the measurements from Dole
et al. (2004) have a slope of --2.0.  On the other hand, the
phenomenological galaxy population model of Lagache et al. (2004),
which has been updated to be consistent with the more recent MIPS
results, is similar in its fundamentals, i.e.  starburst galaxies make
up the bulk of the contribution to CFIB fluctuations and passively
evolving spirals contribute most of the remainder.  The power spectrum
prediction for this updated model is not given, however, so we do not
know how this compares to our results.  Future work, including the
following, would be a clear way forward to finding the point of
model-data discrepancy, and further refinements in our understanding:
(i) calculation of the power spectrum with the newest model
(incorporating Spitzer data), (ii) testing the model source
distribution against the actual source distributions in the deepest
MIPS far-IR surveys, for each identified source type, and (iii) using
those identified source types to determine relative bias (both
directly with the identified sources within the survey, and indirectly
by associating a source type with the color classifications in optical
surveys).  With information on source distributions to a much lower
flux, with a power spectrum calculation based on an updated model,
with detailed comparisons to optical data, we will be able to better
constrain the galaxy population models; adding to all these
constraints the structure measurements herein will then realize the
opportunity to more precisely measure the structure.

The study of structure in the distribution of IR galaxies has
important potential for illuminating the processes that have taken
place since the CMB era.  A valuable feature of these far-IR
measurements is their virtual immunity to extinction and any
extinction-related bias (unlike optical galaxy structure
measurements).  Additionally, current models suggest that the source
population that contributes the majority of the power spectrum
structure signal is very simple - only starbursts and IR-bright
spirals contribute significantly.  If these models are correct, this
would also reduce problems of source-type bias that are present in
some optical studies.  Continued refinement of these measurements, and
our knowledge of the source populations, will yield a useful new view
of structure very much complimentary to that given by optical
measurements.

\subsection*{5.2 Possibilities for Future Improvements}


We have corrected most of the scan-pattern related structure, and we
have shown that the residual scan-pattern related structure in the
power spectrum is small compared to the structure we have measured.
However, some residual scan-pattern related structure is still present
in the map, and this is the first item we would like to improve on in
future work. In terms of data reduction, various frequency space
filtering methods for removing stripes seem promising, and have been
demonstrated on other types of maps (see, e.g., Miville-Desch\^{e}nes
\& Lagache \cite{Deschenes05}).  In addition, various algorithms are
available for optimum statistical weighting of map data to reduce map
deviation.  We found, however, that unless the deviation in the maps
are significantly reduced, these algorithms do not produce good
results on these data.


We find that the greatest potential for improving these measurements
is in acquiring new data on the SLH field in a manner appropriate for
background observations. The MIPS observations are truly exceptional
compared to typical background observations in the degree to which
``cross-linking'' scans were \textit{avoided}.  Virtually all cosmic
microwave/mm background experiments incorporate cross-linking in their
scanning strategy. Such a strategy causes each sky pixel to be
re-sampled along significantly different scanning paths on the sky. A
simplified example of cross-linking scans would be a series of two
rectilinear raster maps of a region oriented at 90\ensuremath{^\circ}
to each other. Two sky positions measured along the same scan across
the first map would be measured on different scans in the second
map. Comparison of repeated observations on the same and different
scans allows identification and measurement of any systematics that
affect the measurements differently on the same and different
scans. In the general case, this technique allows inter-comparison of
measurements made close together in time (i.e. on the same scan) with
those made at much longer time scales. In our case, this comparison
would more clearly identify and more accurately measure the zero-point
offsets. In all of the large MIPS surveys, rectangular regions of the
sky were observed during each AOR, and then immediately repeated on
almost precisely the same path (with only minor exceptions). The
rectangular regions were oriented very nearly parallel, no scans were
made along significantly different directions (except in the very
limited verification observations), and these regions had only small
edge overlap. Cross-linking was essentially \textit{minimized} in the
existing surveys, permitting inter-comparison among only a small
fraction of data measured on different paths in different AORs. It
seems clear that if cross-linking MIPS scans were added to existing
Spitzer survey regions, significant improvement in the background
fluctuation measurement would result.  We proposed a program of
10 hours of MIPS observations to make cross-linking scans on this same
field during Spitzer Cycle 3. We are confident
that the resulting improvement in systematic error will lead to
significantly smaller errors in the low-\textit{k} region of interest,
putting even more pressure on the galaxy population models, and
contributing to our measurement of structure in the universe as traced
by far-IR emitting galaxies.  Ultimately we intend that these improved
measurements will contribute to guiding our understanding of the
physics of formation of structure in galaxies, allowing us to
understand the behavior of luminous matter from the CMB era to today.

\section*{Conclusions}

In this paper we presented co-added maps from two large Spitzer survey
fields observed with the 160 \ensuremath{\mu}m MIPS array.
Instrumental artifacts, and artifacts related to the scan pattern
(i.e. stripes) were observed, but these effects were controlled in two
ways: First, the artifacts were substantially reduced by numerically
calculated corrections.  Second, we carefully measured the errors due
to these artifacts by simulation: We added artifacts, at the same
intensity measured in the real data, to simulated timeline data, and
after the same reduction as for the real data, we compared the
resulting power spectra to that expected. In the end, the errors
introduced into our CFIB power spectrum were shown to have no
significant effects on our conclusions.  We measured a cirrus power
law slope of --3.11 \ensuremath{\pm} 0.30 in the SLH field.
Subtracting this power law yielded a CFIB spectrum dropping rapidly
from \textit{k} \ensuremath{\sim} 0.03 arc min$^{-1}$, to \textit{k} =
0.1 - 0.2 arc min$^{-1}$, and a flatter region at higher \textit{k}.
Any assumption of a power law cirrus component plus a flat power
spectrum from the ensemble of sources, i.e. from a random distribution
of sources, is inconsistent with our observations.  Our results are
consistent with the general characteristics of predictions of a
clustering ``signature'' in the CFIB power spectrum, but are steeper
at low-\textit{k} than some predictions.  This is the first reported
measurement of clustering derived from far-IR (50-200 ${\mu}$m)
observations.

\begin{acknowledgements}
  The authors wish to thank the staff and students of the Institute
  d'Astrophysique Spatiale (IAS), especially Guilaine Lagache,
  Fran\c{c}ois Boulanger, and Herv\'{e} Dole, for their kindness and
  collaboration during and after Grossan's time at the IAS. We wish to
  also thank the staff of the Spitzer SSC, especially at the Help
  Desk, the staff of Eureka Scientific, and the LBNL Institute for
  Nuclear and Particle Astrophysics. We also thank the referee for their 
  careful reading of this paper and valuable suggestions.  
  This work is based on archival
  data obtained with the Spitzer Space Telescope, which is operated by
  the Jet Propulsion Laboratory, California Institute of Technology
  under a contract with NASA. Support for this work was provided by an
  award issued by JPL/Caltech, a Spitzer Archival Research grant, NASA
  grant 1263806.
\end{acknowledgements}


\begin{thebibliography}{}

\bibitem[1999]{Abergel}Abergel A., Andr P., Bacmann A., et al., 1999, in: The Universe as seen by ISO. ESA-SP 427
\bibitem[1996]{Bertin}Bertin, E., \& Arnouts, S. 1996, A\&AS, 117, 393
\bibitem[2002]{Connolly}Connolley, A. J,. et al. 2002, ApJ 579, 42
\bibitem[2004]{Doleetal04}Dole, H., et al. 2004, ApJS 154, 87
\bibitem[2003]{DLP03}Dole, H., Lagache, G., \& Puget, J.-L. 2003, ApJ 585,617
\bibitem[1998]{Falgarone}Falgarone E., 1998, in: Guiderdoni B., Kembhavi A. (eds.) Starbursts: Triggers, Nature and Evolution. Les Houches School
\bibitem[2006]{Frayer}Frayer, D. T., et al. 2006, AJ 131, 250
\bibitem[1992]{Gautier}Gautier T.N.III, Boulanger F., Perault M., Puget J.L., 1992, 
AJ 103, 1313 
\bibitem[2005]{Gordon}Gordon, K. et al. 2005, PASP 117,503
\bibitem[1998]{Herbstmeir}Herbstmeier U., Abraham P., Lemke D., et al., 1998, A\&A 332, 
739
\bibitem[1996]{Kogut}Kogut A., Banday A. J., Bennett C. L., et al., 1996, ApJ 460, 1
\bibitem[2000]{LagachePuget00}Lagache, G. \& Puget, J.-L. 2000, A\&A 357, 5
\bibitem[2000]{Lagacheetal00}Lagache, G., et al. 2000, ``The Extragalactic Background and Its Fluctuations in the Far-Infrared Wavelengths'', in ``ISO Surveys of a Dusty Universe'', Proceedings Ringberg Workshop, (Eds.) D. Lemke, M. Stickel, K. Wilke,  Lecture Notes in Physics 548 (2000)
\bibitem[2001]{LagacheandDole}Lagache, G. \& Dole, H. 2001, A\&A 372, 702
\bibitem[2003]{LDP03}Lagache, G., Dole, H., \& Puget, J.-L. 2003, MNRAS 338, 555
\bibitem[2004]{Lagacheetal04}Lagache, G., et al. 2004, ApJS 154, 112
\bibitem[2000]{Matsuhara}Matsuhara, H., et al. 2000, A\&A 361,407
\bibitem[2005]{Deschenes05}Miville-Desch\^{e}nes, M. A., \& Lagache,
  G., 2005, ApJS, 157, 302 
\bibitem[2002]{Deschenes02}Miville-Desch\^{e}nes, M. A., Lagache, G., Puget, J.-L. 2002, A\&A, 393, 749
\bibitem[2006]{Noriega-Crespo}Noriega-Crespo, A., 2006, August 10, private email communication.
\bibitem[2003]{Perrotta}Perrotta, F. et al. 2003, MNRAS 338,623
\bibitem[2004]{Rieke}Rieke, G., et al. 2004, ApJS 154, 25
\bibitem[1998]{Schlegel}Schlegel D. J., Finkbeiner D. P., Davis M., 1998, ApJ 500, 525
\bibitem[1998]{Wright}Wright E.L., 1998, ApJ 496, 1

\end{thebibliography}
\end{document}